\documentclass[prb,11pt]{revtex4-1}
\usepackage{amsmath}    
\usepackage{color}
\begin{document}%

\title{Thermo-electromagnetic transport}
\author{A. I. Arbab }
\affiliation{Department of Physics, College of Science, Qassim University, P.O. Box 6644, 51452 Buraidah,  KSA}

\begin{abstract}
Maxwell's equations incorporating thermoelectric and thermomagnetic effects are studied. Energy transport involving electric field only flows along the velocity direction and a direction perpendicular to it.  Magnetic energy transport associated with magnetic field only is found to flow along the velocity direction and a direction normal to it.   The temperature is transmitted like an electromagnetic wave traveling at the speed of light. Thermoelectric and thermomagnetic polarizations are induced in  the medium that are directly proportional to the  temperature. An electronic transport due to temperature variations only and without electric or magnetic field is found to be accessible. Electric and magnetic fields due to temperature gradient are shown to arise provided the photon is massive.
\end{abstract}

\maketitle
\section{\textcolor[rgb]{0.00,0.07,1.00}{Introduction}}
Maxwell's equations showed that light is a kind of an electromagnetic field. The electromagnetic field satisfies the wave equation that travels at the speed of light in vacuum \textcolor[rgb]{0.00,0.07,1.00}{\cite{griff}}. The electromagnetic wave has a transverse nature in vacuum, where the electric and magnetic fields direction are always perpendicular to the direction of propagation of the wave. No electromagnetic wave will be emitted if its electric or magnetic field is zero.

We recently formulated Maxwellian equations governing the electric and magnetic fields associated with moving electrons \textcolor[rgb]{0.00,0.07,1.00}{\cite{spinor}}. Owing to this formalism, the electric and magnetic fields are modified to incorporate the electron velocity, in addition to new electric  and magnetic scalars. The electric and magnetic fields besides satisfying Maxwell's equations, they also satisfy Dirac equation in its quaternionic form. Thus,  these fields are those fields which are associated with matter wave inscribed in electrons.
However, when Maxwell's equations are modified, by introducing electric and magnetic scalars that are related to temperature, thermoelectric and thermomagnetic properties are arised. Moreover, new features for the electromagnetic fields are shown to be  allowed. According, the temperature is found to satisfy the wave equation traveling at the speed of light rather than satisfying  the diffusion equation.

We are interested here to study the energy and momentum transported by electrons when (i) only electric field is present, and (ii) only magnetic field is involved, (iii) no electric or magnetic field is present. We are dealing here with the transport of matter waves associated with electrons. In each case we derive the energy and momentum conservation equations, where the energy and momentum are found to flow along the velocity direction, and a perpendicular to it. The momentum density and the stress tensor show that the electrons move like  a fluid with electric and magnetic polarizations that are proportional to the temperature and electron velocity. Moreover, these exits a case when the electric and magnetic fields are induced from temperature gradient.

\section{\textcolor[rgb]{0.00,0.07,1.00}{Modified Maxwell's equations }}

Note that electrons matter wave is described by Dirac equation. Maxwell's equations deal with the evolution of the electromagnetic fields. We introduced a formalism that uses the electromagnetic field as a spinor field (Dirac matter wave), where we have found that the electromagnetic field due to moving electrons obey Maxwell's equations as well, but with additional terms \textcolor[rgb]{0.00,0.07,1.00}{\cite{spinor}}. We thus obtained a set of modified Maxwell's equations that can be seen as a minimal modification. In this formalism Maxwell's equations incorporate two scalars that are coupled to the electromagnetic field as \textcolor[rgb]{0.00,0.07,1.00}{\cite{spinor}}
\begin{equation}
\vec{\nabla}\cdot\vec{E}_D=\frac{\partial\Phi}{\partial t}\,,
\end{equation}
\begin{equation}
\vec{\nabla}\cdot\vec{B}_D=\frac{1}{c^2}\frac{\partial\Psi}{\partial t}-\frac{\beta mc}{\hbar}\,\Lambda\,,
\end{equation}
\begin{equation}
\vec{\nabla}\times\vec{E}_D=-\frac{\partial\vec{B}_D}{\partial t}+\vec{\nabla}\Psi+\frac{\beta mc}{\hbar}\vec{E}\,,
\end{equation}
\begin{equation}
\vec{\nabla}\times\vec{B}_D=\frac{1}{c^2}\frac{\partial\vec{E}_D}{\partial t}+\frac{\beta mc}{\hbar}\,\vec{B}-\vec{\nabla}\Phi\,,
\end{equation}
where
\begin{equation}
\vec{E}_D=\vec{E}+\vec{v}\times\vec{B}+\vec{v}\,\Lambda\,,\qquad\qquad
\vec{B}_D=\vec{B}-\frac{\vec{v}}{c^2}\times\vec{E}\,,
\end{equation}
and that
\begin{equation}
\Psi=\vec{v}\cdot\vec{B}\,,\qquad\qquad
\Phi=\Lambda+\frac{\vec{v}}{c^2}\cdot\vec{E}\,,
\end{equation}
and
 $\beta=\left(\begin{array}{cc}
1 & 0 \\
0 & -1\end{array}\right)$ is a $4\times 4$ Dirac's matrix, and $m$ is the photon mass. The electric and magnetic fields in Eq.(5) are those associated with the matter wave of the electrons.
The energy conservation equation for the system described by Eqs.(1) - (6) is
\begin{equation}
\frac{\partial u_D}{\partial t}+\vec{\nabla}\cdot\vec{S}_D=-\frac{\beta mc\Lambda}{\mu_0\hbar}\,\vec{v}\cdot\vec{B}\,,
\end{equation}
where
\begin{equation}
\vec{S}_D=\mu_0^{-1}\left(\vec{E}_D\times\vec{B}_D-\Phi\vec{E}_D-\Psi\vec{B}_D\right)\,, \qquad u_D=\frac{1}{2\mu_0}\left(B^2+\Phi^2\right)+\frac{\varepsilon_0}{2}\left(E^2_D+\Psi^2\right).
\end{equation}
Using Eq.(6), Eq.(7) can be expressed as
\begin{equation}
\frac{\partial u_D}{\partial t}+\vec{\nabla}\cdot\vec{S}_D=-\frac{\beta mc}{\mu_0\hbar}\,\Lambda\,\Psi\,.
\end{equation}
It is pertinent to mention that the right hand-side of Eq.(9), in the standard electrodynamics energy conservation equation, is $-\vec{J}\cdot\vec{E}$ \textcolor[rgb]{0.00,0.07,1.00}{\cite{griff}}, which vanishes in vacuum. It is interesting to see that the photon mass doesn't affect the energy and momentum densities but  appears as a dissipation power only.
For massless photon, \emph{i.e}., $m=0$, Eqs.(1) - (4) reduce to
\begin{equation}
\vec{\nabla}\cdot\vec{E}_D=\frac{\partial\Phi}{\partial t}\,, \qquad\qquad
\vec{\nabla}\cdot\vec{B}_D=\frac{1}{c^2}\frac{\partial\Psi}{\partial t}\,,
\end{equation}
\begin{equation}
\vec{\nabla}\times\vec{E}_D=-\frac{\partial\vec{B}_D}{\partial t}+\vec{\nabla}\Psi\,,\qquad\qquad
\vec{\nabla}\times\vec{B}_D=\frac{1}{c^2}\frac{\partial\vec{E}_D}{\partial t}-\vec{\nabla}\Phi\,.
\end{equation}
Manipulation of Eqs.(10) and (11),  yields the following wave equations
\begin{equation}
\frac{1}{c^2}\frac{\partial^2\vec{E}_D}{\partial t^2}-\nabla^2\vec{E}_D=0\,,\qquad\qquad
\frac{1}{c^2}\frac{\partial^2\vec{B}_D}{\partial t^2}-\nabla^2\vec{B}_D=0\,.
\end{equation}
The two scalars $\Phi$ and $\Psi$ also satisfy the wave equations
\begin{equation}
\frac{1}{c^2}\frac{\partial^2\Phi}{\partial t^2}-\nabla^2\Phi=0\,,
\qquad\qquad
\frac{1}{c^2}\frac{\partial^2\Psi}{\partial t^2}-\nabla^2\Psi=0\,.
\end{equation}

\section{\textcolor[rgb]{0.00,0.07,1.00}{Energy and momentum conservation equations}}
For massless photon and $\Lambda=0$, Eqs.(8) and (9) become
\begin{equation}
\frac{\partial u_D}{\partial t}+\vec{\nabla}\cdot\vec{S}_D=0\,,
\end{equation}
where
\begin{equation}
\vec{S}_D=\mu_0^{-1}\left(\vec{E}_D\times\vec{B}_D-\Phi\vec{E}_D-\Psi\vec{B}_D\right)\,, \qquad u_D=\frac{1}{2\mu_0}\left(B^2_D+\Phi^2\right)+\frac{\varepsilon_0}{2}\left(E^2_D+\Psi^2\right).
\end{equation}
This equation states that the energy is carried by the combined electromagnetic fields, $\vec{E}_D$ and $\vec{B}_D$. The energy flows without loss (dissipation) along a direction perpendicular to the fields, $\vec{E}_D$ and $\vec{B}_D$, and along a plane containing these two fields. It is apparent from Eq.(15) that $\Phi$ behaves like  a magnetic scalar, and $\Psi$  as an electric scalar that are coupled to the electric and magnetic fields, respectively.

To obtain the momentum conservation equation, we cross multiply the first equation in Eq.(11) by $\vec{E}_D$, and the second equation in Eq.(11) by $\vec{B}_D$ and add the two resulting equations, using the vector identity, $$\vec{\nabla}\left(\frac{A^2}{2}\right)=\vec{A}\times(\vec{\nabla}\times\vec{A})+(\vec{A}\cdot\vec{\nabla})\vec{A},$$
 where we will obtain
\begin{equation}
-\frac{\partial \vec{g}_D}{\partial t}+\vec{\nabla}\cdot\vec{\sigma}=\vec{\nabla}\times(\varepsilon_0\vec{E}\Psi-\mu_0^{-1}\vec{B}\Phi)\,,
\end{equation}
where
\begin{equation}
\vec{g}_D=\varepsilon_0\left(\vec{E}_D\times\vec{B}_D+\vec{E}_D\Phi+\vec{B}_D\Psi\right)\,,\qquad \sigma_{ij}=\varepsilon_0E_{Di}E_{Dj}-\frac{\varepsilon_0}{2}\,\delta_{ij}(E^2_D-\Psi^2)+\mu_0^{-1}B_{Di}B_{Dj}-\frac{1}{2\mu_0}\delta_{ij}(B^2_D-\Phi^2)
\end{equation}
It is interesting to note from Eqs.(15) and (17)  that the relation $\vec{g}_D=\frac{\vec{S}_D}{c^2}$ is no longer valid, except when $\Phi=\Psi=0$.
As evident from Eq.(16), a force density can be defined as
\begin{equation}
\vec{f}_D=\vec{\nabla}\times(\varepsilon_0\vec{E}\Psi-\mu_0^{-1}\vec{B}\Phi).
\end{equation}
And  since the torque and force densities are connected by the relation $\vec{f}=\frac{1}{2}\,(\vec{\nabla}\times\vec{\tau}\,)$ \textcolor[rgb]{0.00,0.07,1.00}{\cite{spin1}}, then the torque density induced by the matter field will be
\begin{equation}
\vec{\tau}_D=2\,(\varepsilon_0\vec{E}\Psi-\mu_0^{-1}\vec{B}\Phi).
\end{equation}
 The above torque requires a priori the presence of the two scalars, $\Psi$ and $\Phi$, as defined in Eq.(6). The scalar $\Psi$ requires the quantum particle to be moving, but $\Phi$ acts even for stationary particle. Using Eq.(6), the above torque becomes
\begin{equation}
\vec{\tau}_D=2\,\varepsilon_0(\vec{v}\cdot\vec{B})\,\vec{E}-2\varepsilon_0(\vec{v}\cdot\vec{E})\,\vec{B}.
\end{equation}
This shows that the  torque acts one the plane made by the electric  and magnetic fields directions. A similar torque is found to show up when considering the electromagnetic field as a fluid \textcolor[rgb]{0.00,0.07,1.00}{\cite{fluid}}.

If the photon is massive, $m\ne 0$, then  Eq.(3) suggests an electric field
 $$\hspace{3cm} \vec{E}=\pm\frac{\hbar}{mc}\vec{\nabla}\Psi\,.\hspace{3cm} (a)$$
and upon employing the ansatz  that $\Psi=\xi T$, where $\xi$ is some constant, Eq.(a) becomes
$$\hspace{3cm} \vec{E}_T=\pm\,\frac{\xi\hbar}{mc}\vec{\nabla} T\,.\hspace{3cm} (b)$$
This is an interesting equation stating that an electric field due to temperature gradient is created. One can associate a potential with the above electric field as $V_T=\frac{\xi\hbar}{mc}\, T$. Hence, a potential difference will develop whenever a temperature difference occurs, \emph{i.e}, $\Delta V=\frac{\xi\hbar}{mc}\,\Delta T$. This suggests a quantum Seebeck coefficient, $S_q=\frac{\xi\hbar}{mc}$, relating the two variations. The photon mass in a conductor is given by $m=\frac{1}{2}\,\mu_0\sigma\,\hbar$ \textcolor[rgb]{0.00,0.07,1.00}{\cite{analogy}}. Therefore, one can write, $S_q=2\xi/(\mu_0\sigma c)$.  Note that in Seebeck effect  a voltage due to temperature difference is created, but here an electric field is created due to the temperature gradient \textcolor[rgb]{0.00,0.07,1.00}{\cite{seebeck}}. This electric field would be important in the nanoscale wirings. Equations (10) and  (11) can be seen as Maxwell's equations for non-isothermal medium.

Let us now assume that $\Lambda=0$ but $m\ne0$ and express the scalar $\Phi$ as, $\Phi=\frac{\vec{v}}{c^2}\cdot\vec{E}=\eta\, T$ so that Eq.(4) suggests a magnetic field
 $$\hspace{3cm} \vec{B}=\pm\frac{\hbar}{mc}\,\vec{\nabla}\Phi\,,\hspace{3cm} (c)$$
which can be expressed as
$$\hspace{3cm} \vec{B}_T=\pm\,\frac{\eta\,\hbar}{mc}\,\vec{\nabla} T\,.\hspace{3cm} (d)$$
This is an interesting equation stating that a magnetic field due to temperature gradient is created. This magnetic field is of quantum nature. One can associate a magnetic potential with the above magnetic field as, $V_B=\frac{\eta\,\hbar}{mc}\,T$, so that its gradient gives rise to a magnetic field. A thermal magnetic field  is recently claimed to be generated from temperature gradient that is referred to as the Magnetic Seebeck effect \textcolor[rgb]{0.00,0.07,1.00}{\cite{mag}}.

\section{\textcolor[rgb]{0.00,0.07,1.00}{Electronic transport with  electric field only }}
It is now possible to have an electronic transport with pure electric field with vanishing magnetic field, $\vec{B}=0$. This is allowed because of the presence of the two scalars, $\Phi$ and $\Psi$ that are coupled to the electric and magnetic fields. Owing to Eqs.(15) and (17), this wave carries energy and momentum given by
\begin{equation}
\vec{S}_D=-(\rho_Ec^2)\,\vec{v}\,, \qquad \qquad u_D=\frac{\varepsilon_0}{2}E^2+\frac{1}{2}\,\rho_Ev^2\,,\qquad\qquad \rho_E=\frac{\varepsilon_0}{c^2}\,E^2\,,
\end{equation}
and
\begin{equation}
\vec{g}=2\varepsilon_0\left(\frac{\vec{v}}{c^2}\cdot\vec{E}\right)\vec{E}-\rho_E\vec{v}\,,\qquad \sigma_{ij}=-\rho_E\,v_iv_j-\frac{1}{2}\,\rho_E c^2\,\delta_{ij}+\varepsilon_0(1-\frac{v^2}{c^2})\,E_iE_j+\frac{\varepsilon_0}{c^2}(v_iE_j+v_jE_i)(\vec{v}\cdot\vec{E})
\end{equation}
using Eqs.(5) and (6). Notice that the above stress tensor is a symmetric tensor. Moreover, the mass density energy is directly proportional to the magnetic energy density. Thus, when no magnetic field is present, the total energy density is equal to the electric energy density plus the mass energy density, that is contained in the electric field, of the moving electrons. Notice that while the energy flows along the velocity direction, the momentum spreads along the velocity and electric field directions.

If we now relate the scalar gradient to an additional current (\emph{thermal electric current}) that arises due to temperature difference, then Eqs.(10) and (11) suggest that
\begin{equation}
\mu_0\vec{J}_T=-\vec{\nabla}\Phi\,,\qquad \qquad \rho_T=\varepsilon_0\frac{\partial\Phi}{\partial t}\,.
\end{equation}

Let us now apply the ansatz $\Phi=\eta\, T$ in Eqs.(21) and (22), employing Eq.(6), to obtain
\begin{equation}
\vec{S}_D=-(\rho_Ec^2)\,\vec{v}\,, \qquad \qquad u_D=\frac{\varepsilon_0}{2}E^2+\frac{1}{2}\,\rho_Ev^2\,,\qquad\qquad \rho_E=\frac{\varepsilon_0}{c^2}\,E^2\,,
\end{equation}
 \begin{equation}
\vec{g}=2\varepsilon_0\eta\, T\vec{E}-\rho_E\vec{v}\,,\qquad \sigma_{ij}=-\rho_E\,v_iv_j-\frac{1}{2}\,\rho_E c^2\,\delta_{ij}+\varepsilon_0\left(1-\frac{v^2}{c^2}\right)E_iE_j+\varepsilon_0\eta\, T(v_iE_j+v_jE_i).
\end{equation}
Equation (25) states that some thermal stress is induced due to the propagation of temperature  inside the material that causes the current to flow. The momentum density acts over the $\vec{v}-\vec{E}$ plane, and therefore  transverse current and temperature difference may occur. Remarkably, Eqs.(24) and (25) govern the transport of electrons inside the material in presence of electric field only. Note that in electromagnetism, the transport of electromagnetic wave (energy) requires  a priori the presence of dynamical electric and magnetic fields.

In the standard electromagnetic theory, the momentum carried by the electrons is always along the electric field direction. This is so when the first term in the first equation in Eq.(25) vanishes.  In fact, a voltage occurs across the transverse direction in aconductor/semicondutor is called the Hall voltage. Similarly, a transverse temperature difference occurs when a longitudinal one is applied that is know as the Nernst effect \textcolor[rgb]{0.00,0.07,1.00}{\cite{nernst}}.  The thermoelectric and magnetic effects pertaining to Eq.(23) are studied in \textcolor[rgb]{0.00,0.07,1.00}{{\cite{thermo}}}. The last term in Eq.(25) can be expressed as
\begin{equation}
\sigma_{ij}^P=P_iE_j+P_jE_i\,,\qquad\qquad P_i=\varepsilon_0\eta\, Tv_i\,,
\end{equation}
where $\vec{P}$ is the polarization vector. The torque on the electric dipole is related to the polarization by the relation, $\vec{\tau}=\vec{P}\times\vec{E}$. It is interesting to see that the electric polarization is now temperature and velocity dependent, \emph{i.e}., $P\propto T$ and $P\propto v$. Moreover, one can define a thermal electric field as arising from the electron motion at a given temperature, as $\vec{E}_T=\eta\, T\vec{v}$\,. This electric field can give rise to an electric current. Furthermore, one can define an electric mobility associated with this thermal electric field as, $\mu=1/(\eta\, T)$. In fact, due to phonon scattering in the material, the electron mobility decreases with increasing temperature.

An educated guess for $\eta$, \emph{i.e.},  is $\eta=\mu_0k_B\sigma/e$ yields a mobility $\mu=e/(\mu_0\sigma k_B\, T)$, where $\sigma$ is the electric conductivity and $k_B$ is the Boltzmann's constant  \textcolor[rgb]{0.00,0.07,1.00}{\cite{thermo}}. Recall that the electric mobility of the material is given by $\mu=e\,\tau/m_e=\sigma/(ne)$, where $\tau$ is the scattering time, and $m_e$ is the electron mass. Equating the two mobilities yields a critical conductivity, $\sigma_0=m_e/(\mu_0k_BT\tau)$. However, we have seen that a quantum conductivity can be established by the relation, $\sigma=2m/(\mu_0\hbar)$, where $m$ is the photon mass \textcolor[rgb]{0.00,0.07,1.00}{\cite{analogy}}. Equating the two conductivities for the photon yields the uncertainty - like relation, $k_BT\,\tau=\hbar/2$. This implies that the relaxation time is $\tau=\hbar/(2k_BT)$.

Now Eq.(23) suggest that the temperature satisfies the wave equation
\begin{equation}
\frac{1}{c^2}\frac{\partial^2T}{\partial t^2}-\nabla^2T=0.
\end{equation}
It is interesting that like the electric field that propagates at the speed of light, the temperature too propagates at the speed of light. Moreover, the temperature couples to the electric field as evident from Eq.(25). It is also evident from Eq.(20) that no torque can be applied on electrons as far as no magnetic field is present.

\section{\textcolor[rgb]{0.00,0.07,1.00}{Electronic transport with magnetic field only }}
It is now possible to have an electronic transport with pure magnetic field, and without electric field, $\vec{E}=0$. This is allowed because of the presence of the two scalars, $\Phi$ and $\Psi$ that are coupled to the electric field and magnetic field. Owing to Eqs.(15) and (17), this wave carries energy and momentum given by
\begin{equation}
\vec{S}_D=\mu_0^{-1}(\vec{v}\cdot\vec{B})\vec{B}-\rho_Bc^2\vec{v}\,\,, \qquad u_D=\frac{B^2}{2\mu_0}+\frac{1}{2}\,\rho_Bv^2\,,\qquad \rho_B=\varepsilon_0B^2\,,
\end{equation}
and
\begin{equation}
\vec{g}=2\varepsilon_0(\vec{v}\cdot\vec{B})\vec{B}-\rho_B\vec{v}\,,\,\,\, \sigma_{ij}=-\rho_Bv_iv_j-\frac{1}{2}\,\rho_B\,c^2\left(1-\frac{v^2}{c^2}\right)\,\delta_{ij}+\frac{1}{\mu_0}\left(1-\frac{v^2}{c^2}\right)B_iB_j+\varepsilon_0(v_iB_j+v_jB_i)(\vec{v}\cdot\vec{B})
\end{equation}
Thus, when no electric field is present, the total energy density is equal to the magnetic energy density plus the mass energy density, that is contained in the magnetic field, of the moving electrons.

Let us now assume that the scalar $\Psi$ is related to temperature by the relation, $\Psi=\xi\, T$, where $\xi$ is some constant, apply this in Eqs.(10), (11) and (6),  to obtain
\begin{equation}
\vec{J}_m=-\vec{\nabla}\Psi\,,\qquad\qquad \rho_m=\frac{1}{c^2}\frac{\partial\Psi}{\partial t}\,,
\end{equation}
where $\rho_m$ and $\vec{J}_m$ are the magnetic charge and magnetic current densities. Recall that $\Psi=\xi\, T$ so that the spatial and temporal temperature variations can induce magnetic charge and magnetic current densities.

Now Eqs.(28) and (29) become
\begin{equation}
\vec{S}_D=\mu_0^{-1}\xi\, T\vec{B}-\rho_Bc^2\vec{v}\,\,, \qquad u_D=\frac{B^2}{2\mu_0}+\frac{1}{2}\,\rho_Bv^2\,,\qquad \rho_B=\varepsilon_0B^2\,,
\end{equation}
and
\begin{equation}
\vec{g}=2\varepsilon_0\,\xi\,T\vec{B}-\rho_B\vec{v}\,,\qquad \sigma_{ij}=-\rho_Bv_iv_j-\frac{1}{2}\,\rho_B\,c^2\left(1-\frac{v^2}{c^2}\right)\,\delta_{ij}+\frac{1}{\mu_0}\left(1-\frac{v^2}{c^2}\right)B_iB_j+\varepsilon_0\xi\,T(v_iB_j+v_jB_i).
\end{equation}
For low electron velocities, the stress tensor in Eq.(33) reduces to
\begin{equation}
 \sigma_{ij}=-\rho_Bv_iv_j+\varepsilon_0\xi\,T(v_iB_j+v_jB_i)+\frac{B_iB_j}{\mu_0}-\frac{B^2}{2\mu_0}\,\delta_{ij}\,.
\end{equation}
The first term in Eq.(34) is the stress due to electrons moving as a fluid, the second term is a coupling between electrons momentum, temperature and magnetic field, the last term expresses the stress due to the magnetic field alone.
The the second term in the stress tensor  in Eq.(34) can be casted in the form
\begin{equation}
 \sigma^M_{ij}=M_iB_j+M_jB_i\,,\qquad\qquad M_i=\varepsilon_0\xi\,Tv_i\,,
\end{equation}
where $\vec{M}$ is the magnetization vector, and that it is proportional to the temperature, $T$. The torque on the magnetic dipole is related to the polarization by the relation, $\vec{\tau}=\vec{M}\times\vec{B}$, or $\tau_i=-\epsilon_{ijk}\sigma_{jk}$. It is interesting here to associate a thermal magnetic field, that is directly related to the electrons velocity, at a given temperature by $\vec{B}_T=\frac{\xi}{c^2}\,T\vec{v}$\,. Furthermore, one can define a magnetic mobility associated with this thermal magnetic field as, $\mu_M=c/(\xi\, T)$.

\section{\textcolor[rgb]{0.00,0.07,1.00}{Electronic transport without an electromagnetic field }}
We would like to study here the transport of the electrons when no electric or magnetic field is present, but we allow here $\Lambda$ not to be zero. In this case one substitute $\vec{E}=0$ and $\vec{B}=0$ in Eqs.(15) and (17), to obtain
\begin{equation}
\vec{S}_D=-\frac{\Phi^2}{\mu_0}\,\vec{v}\,, \qquad \qquad u_D=\frac{\Phi^2}{2\mu_0}\left(1+\frac{v^2}{c^2}\right)\,,
\end{equation}
and
\begin{equation}
\vec{g}_D=\varepsilon_0\Phi^2\,\vec{v}\,,\qquad \sigma_{ij}=\varepsilon_0\Phi^2v_iv_j+\frac{\Phi^2}{2\mu_0}\left(1-\frac{v^2}{c^2}\right)\delta_{ij}\,.
\end{equation}
And as done before, one can relate  the scalar, $\Phi=\Lambda$, to the temperature, $T$. Therefore, even no electric or magnetic field is present electrons in the material can be transported carrying energy and momentum given by Eqs.(36) and (37). Now here no  energy or momentum is spread in a plane as before. Energy and momentum are  transported totally along the electron velocity direction. Thus, electrons carry  heat and disperse it along with it. For relatively low velocities,  Eqs.(36) and (37) reduce to
\begin{equation}
\vec{S}_D=-2u_D\,\vec{v}\,, \qquad \qquad
\vec{g}_D=2\left(\frac{u_D}{c^2}\right)\,\vec{v}\,,\qquad \sigma_{ij}=2\left(\frac{u_D}{c^2}\right)v_iv_j+u_D\delta_{ij}\,,\qquad u_D=\frac{\Phi^2}{2\mu_0}\,,
\end{equation}
where the energy density, $u_D\propto T^2$.  The term $\rho=2u_D/c^2$ can be seen as the mass energy density carried by the electrons.  Hence, Eq.(38) can be written in the form
\begin{equation}
\vec{S}_D=-\rho\,c^2\,\vec{v}\,, \qquad \qquad
\vec{g}_D=\rho\,\vec{v}\,,\qquad \sigma_{ij}=\rho\,v_iv_j+p\,\delta_{ij}\,.
\end{equation}
 The term $p=\frac{1}{2}\,\rho\,c^2$ can be seen as a pressure term, and that electrons move like a perfect fluid.
Note that the force density associated with this case is zero, \emph{i.e.}, $\vec{f}_D=0$, as can be seen from Eq.(18).

\section{\textcolor[rgb]{0.00,0.07,1.00}{Concluding Remarks}}
Using the modified Maxwell's equations we recently introduced, that generalize the electric and magnetic fields due to moving electron, we have found that different electronic transports are possible. By relating the electric and magnetic scalars appearing in the modified Maxwell's equations to temperature many interesting cases are shown to occur. Because of these scalars, the energy and momentum can  be transported by electrons. We have studied here electronic transports with  electric field only,  magnetic field only, and no fields. In these cases the temperature plays the key role. The energy and momentum flow along the velocity direction and a direction perpendicular to it. When no electric or magnetic field is present, the heat (temperature) is found to contribute to the electronic transport where the energy and momentum are directed along the velocity direction only. We have shown that  electric and magnetic fields can be induced in a material when a temperature gradient exists between two  points.


\end{document}